\documentclass[12pt]{article}
\usepackage[english,german,french,polish]{babel}
\usepackage[T1]{fontenc}
\usepackage{amsfonts}

\begin{document}

\selectlanguage{english}

\baselineskip 0.76cm
\topmargin -0.6in
\oddsidemargin -0.1in

\let\ni=\noindent

\renewcommand{\thefootnote}{\fnsymbol{footnote}}

\newcommand{\SM}{Standard Model }

\pagestyle {plain}

\setcounter{page}{1}

\pagestyle{empty}

~~~

\begin{flushright}
IFT-- 07/3
\end{flushright}

\vspace{0.5cm}

{\large\centerline{\bf Towards a realistic neutrino mass formula:}}

{\large\centerline{\bf part two{\footnote{Work supported in part by the Polish Ministry of Higher Education and Science, grant 1 P03B 099 29 (2005-2007).}} {\footnote {This is a sequel of the paper {\tt hep-ph/0609187}.}}}} 

\vspace{0.4cm}

{\centerline {\sc Wojciech Kr\'{o}likowski}}

\vspace{0.3cm}

{\centerline {\it Institute of Theoretical Physics, University of Warsaw}}

{\centerline {\it Ho\.{z}a 69,~~PL--00--681 Warszawa, ~Poland}}

\vspace{0.5cm}

{\centerline{\bf Abstract}}

\vspace{0.2cm}

A two-parameter neutrino mass formula is described, giving a moderately hierarchical spectrum 
$m_1 < m_2 < m_3 $ consistent with the experimental estimates of $\Delta m^2_{21} $ and 
$\Delta m^2_{32}$. The formula follows from a three-parameter empirical neutrino mass 
formula through imposing a parameter constraint and leads to a very simple neutrino mass 
sum rule $m_3 = 18 m_2 - 48 m_1$. Some alternative parameter constraints, used tentatively 
to eliminate one of three parameters, are compared.

\vspace{0.5cm}

\ni PACS numbers: 12.15.Ff , 14.60.Pq  

\vspace{1.6cm}

\ni June 2007  

\vfill\eject

~~~
\pagestyle {plain}

\setcounter{page}{1}

\vspace{0.3cm}

Recently, we have discussed some parameter constraints [1] in an empirical mass formula [2] for three active mass neutrinos $ \nu_1,\nu_2, \nu_3$ related to three active flavor neutrinos $ \nu_e,\nu_\mu, \nu_\tau $ through the unitary mixing transformation $\nu_\alpha = \sum_i U_{\alpha i} \nu_i \;(\alpha = e, \mu, \tau$ and $i = 1,2,3)$. Originally, this neutrino mass formula involves three free parameters $\mu, \varepsilon, \xi$ and gets the form:

\begin{equation}
m_{i}  = \mu \, \rho_i \left[1- \frac{1}{\xi} \left(N^2_i + \frac{\varepsilon -1}{N^2_i}\right)\right]\;(i = 1,2,3) \;,
\end{equation}

\ni where 

\begin{equation}
N_1 = 1 \;,\; N_2 = 3 \;,\; N_3 = 5 \;\;\;,\;\;\; \rho_1 = \frac{1}{29} \;,\; \rho_2 = \frac{4}{29} \;,\; \rho_3 = \frac{24}{29}
\end{equation}

\ni ($\sum_i \rho_i = 1$), the latter three numbers having been called the generation-weighting factors. Here and in Ref [1], $\mu$ denotes the product $\mu \xi/\zeta$ previously appearing in Ref. [2]. Explicitly, the formula (1) can be rewritten as

\begin{eqnarray}
m_1 & = & \frac{\mu}{29} \left(1 - \frac{1}{\xi}\varepsilon\right) \,, \nonumber \\
m_2 & = & \frac{\mu}{29}\, 4\left(1 -\frac{1}{\xi}\frac{80 + \varepsilon}{9}\right) \,, \nonumber \\
m_3 & = & \frac{\mu}{29} \,24\left( 1 -\frac{1}{\xi}\,\frac{624 + \varepsilon}{25}\right) \,.
\end{eqnarray}

This is a specific linear transformation of three free parameters $\mu, \mu\varepsilon/\xi, \mu/\xi $ into three masses $m_1, m_2, m_3 $, giving no mass predictions, {\it unless} the parameters $\mu, \varepsilon, \xi $ are constrained. In fact, in the paper [1], four different parameter constraints

\begin{equation}
\varepsilon/\xi = 1 \;, \varepsilon/\xi = 1- 1/\xi \;, \varepsilon/\xi = 0 \;, 1/\xi = 0 
\end{equation}

\ni were tentatively considered, reducing the mass formula (1) to four two-parameter forms. Then, 
the corresponding neutrino mass spectra were predicted, when the input of experimental estimates for 
$\Delta m^2_{21} $ and $\Delta m^2_{32}$ was used. In general, the transformation inverse to (3) 

\begin{eqnarray}
\mu & = & -29 \frac{125}{4608} \left[m_{3} - \frac{6}{125}(351 m_{2} - 136m_{1})\right] \,, \nonumber \\
\varepsilon & = & 10\, \frac{m_{3} - \frac{6}{125}(351 m_{2} - 904 m_{1}) }{m_{3} - \frac{6}{25}(27 m_{2} - 8m_{1})} \,, \nonumber \\
\xi & = & 10\; \frac{m_{3} - \frac{6}{125}(351 m_{2} - 136 m_{1})}{m_{3} - \frac{6}{25}(27 m_{2} - 8m_{1})}  \, 
\end{eqnarray}

\ni enables us to determine the parameters $\mu, \varepsilon, \xi $, if the masses $m_1, m_2, m_3 $ are known.

In the present note, we discuss still another parameter constraint 

\begin{equation}
\varepsilon =  1 \;\;\;({\rm or}\; \varepsilon/\xi = 1/\xi)\,, 
\end{equation}

\ni while in the first part of this paper, {\tt hep-ph/0609187}, the parameter constraint  $\varepsilon = 0$ (or $\varepsilon/\xi = 0$) was considered. Then, the neutrino mass formula (1) is reduced to the two-parameter form

\begin{equation} 
m_{i} = \mu \rho_i \left(1 - \frac{1}{\xi} N^2_i \right)\;\;(i = 1,2,3)
\end{equation}

\ni or, explicitly,

\begin{eqnarray}
m_1 & = & \frac{\mu}{29} \left(1 - \frac{1}{\xi}\right) \,, \nonumber \\
m_2 & = & \frac{\mu}{29}\, 4\left(1 - 9\frac{1}{\xi}\right) \,, \nonumber \\
m_3 & = & \frac{\mu}{29} \,24\left( 1 - 25\frac{1}{\xi}\right) \,.
\end{eqnarray}

It is easy to see that in the case of Eqs. (7) or (8) the following very simple mass sum rule holds:

\begin{equation}
m_3 = 6(3m_2 - 8 m_1) \;,
\end{equation}

\ni while with the parameter constraint $\varepsilon = 0$ (or $\varepsilon/\xi = 0$) a different mass sum rule $m_3 = (6/125)(351m_2 - 904 m_1)$ worked. In addition, we get here the relations

\begin{equation}
\frac{1}{\xi} = \frac{m_2 - 4 m_1}{m_2 - 36 m_1} \,,\; \mu = 29 \frac{m_1}{1 - 1/\xi} = -29 \frac{m_2 - 36 m_1}{32}
\end{equation}

\ni as a consequence of the general inverse transformation (5) and the mass sum rule (9).

Making use of the mass sum rule (9), $m_3 = 18m_2 - 48 m_1$, and the identity 

\begin{equation}
m^2_3 \equiv \Delta m^2_{32} + m^2_2 \equiv \lambda \Delta m^2_{21} + m^2_2 \equiv (\lambda + 1)m^2_2 - \lambda m^2_1 \,,
\end{equation}

\ni where we have 

\begin{equation} 
 \lambda\equiv \frac{\Delta m^2_{32}}{\Delta m^2_{21}} \sim 30
\end{equation}

\ni due to the experimental estimates $\Delta m^2_{21} \sim 8.0\times 10^{-5}\;{\rm eV}^2$ and $\Delta m^2_{32} \sim 2.4\times 10^{-3}\;{\rm eV}^2$ [3], we obtain for the ratio

\begin{equation}
r \equiv \frac{m_1}{m_2} 
\end{equation}

\ni the quadratic equation

\begin{equation}
(2304 + \lambda) r^2 - 1728 r + 324 - \lambda - 1 = 0 \;.
\end{equation}

\ni Hence, with $\lambda \sim 30$, we get two solutions

\begin{equation}
r \sim \left\{\begin{array}{l} 0.263 = 0.26 \\ 0.477 = 0.48 \end{array} \right. \;.
\end{equation}

\ni Using the ratio $r$, we can evaluate the following neutrino mass spectrum:

\begin{eqnarray}
m_2 & \equiv & \sqrt{\frac{\Delta m^2_{21}}{1- r^2}} \sim \left\{\begin{array}{l} 9.27\times 10^{-3}\;{\rm eV} = 9.3\times 10^{-3}\;{\rm eV} \\ 10.2\times 10^{-3}\;{\rm eV} = 10\times 10^{-3}\;{\rm eV} \end{array} \right. \,, \\
m_1 & \equiv & \;\;r m_2\;\; \sim \left\{\begin{array}{l} 2.44\times 10^{-3}\;{\rm eV} = 2.4\times 10^{-3}\;{\rm eV} \\ 4.86\times 10^{-3}\;{\rm eV} = 4.9\times 10^{-3}\;{\rm eV} \end{array} \right. \,, \\
m_3 & = & 18m_2 - 48 m_1 \sim \left\{\begin{array}{l} \;\;\;4.99\times 10^{-2}\;{\rm eV} = \;\;\;5.0\times 10^{-2}\;{\rm eV} \\ -5.00\times 10^{-2}\;{\rm eV} = -5.0\times 10^{-2}\;{\rm eV} \end{array} \right. \,. 
\end{eqnarray}

\ni  Here, the second solution (15) for $r$ can be rejected, as in this case the masses have different signs (what we exclude). Thus, the predicted neutrino mass spectrum is:

\begin{equation}
m_1 \sim  2.4\times 10^{-3}\;{\rm eV} \;,\; m_2 \sim 9.3\times 10^{-3}\;{\rm eV} \;,\; m_3 \sim  5.0\times 10^{-2}\;{\rm eV} \;,
\end{equation}

\ni when $\Delta m^2_{21} \sim 8.0\times 10^{-5}\;{\rm eV}^2$ and $\Delta m^2_{32} \sim 2.4\times 10^{-3}\;{\rm eV}^2$ are the input. Then, from Eqs. (10) we calculate the parameters

\begin{equation}
\frac{1}{\xi} \sim 6.1\times 10^{-3} \;,\; \mu \sim 7.1\times 10^{-2}\;{\rm eV} \;. 
\end{equation}

This result, valid for the parameter constraint $\varepsilon = 1$ (or $\varepsilon/\xi = 1/\xi$), can be compared in  the following table with our previous results [1] obtained for the alternative parameter constraints (4):

\begin{center}
\begin{tabular}{c|c|c|c|c|c}
$\varepsilon/\xi $ & $1/\xi \,(10^{-3})$ & $\mu\, (10^{-2}\,{\rm eV})$ & $m_1\, (10^{-3}\,{\rm eV})$ & $m_2\, (10^{-3}\,{\rm eV})$ & $m_3\, (10^{-2}\,{\rm eV})$ \\ 
\hline & & & & & \\
1~ & 8.1  & 7.9 & 0~ & 8.9 & 5.0 \\ 
$1- 1/\xi$ & 8.1 & 7.9 & 0.022 & 8.9 & 5.0 \\ 
$1/\xi$ & 6.1 & 7.1 & 2.4 & 9.3 & 5.0 \\ 
0~ & 6.1  & 7.1 & 2.5 & 9.3 & 5.0 \\ 
-8.8~ & 0 & 4.5 & 15\,~~~ & 12\,~~~ & 5.1   
\end{tabular}
\end{center}

\vspace{0.2cm} 

\ni Here, $|\Delta m^2_{21}| \sim 8.0\times 10^{-5}\;{\rm eV}^2$ and $\Delta m^2_{32} \sim 2.4\times 10^{-3}\;{\rm eV}^2$ are used as the input. We can see that the results for $\varepsilon = 0$ (or $\varepsilon/\xi = 0$) and $\varepsilon = 1$ (or $\varepsilon/\xi = 1/\xi$) are similar, since in both cases $1/\xi $ is very small (but it is nonzero, as for $1/\xi \rightarrow 0$ we get $\varepsilon/\xi \rightarrow  -8.8$, when $|\Delta m^2_{21}| \sim 8.0\times 10^{-5}\;{\rm eV}^2$ and $\Delta m^2_{32} \sim 2.4\times 10^{-3}\;{\rm eV}^2$ are the input). 

Mass constraints may sometimes be tried in place of the parameter constraints. For instance, following a recent conjecture [4], the mass ratio $ r \equiv m_1/m_2$ may be related to the so-called golden ratio $\varphi \equiv (1/2)(1 + \sqrt5) = 1.618034$ ($\varphi = 1 + 1/\varphi $) through the formula $r = 1/\varphi^2$. Then,

\begin{equation}
m_1 \sim  3.7\times 10^{-3}\;{\rm eV} \;,\; m_2 \sim 9.7\times 10^{-3}\;{\rm eV} \;,\; m_3 \sim  5.0\times 10^{-2}\;{\rm eV} \;,
\end{equation}

\ni when the experimental input of $\Delta m^2_{21} \sim 8.0\times 10^{-5}\;{\rm eV}^2$ and $\Delta m^2_{32} \sim 2.4\times 10^{-3}\;{\rm eV}^2$ is applied. Hence, due to Eqs. (5),

\begin{equation}
\mu \sim 7.0\times 10^{-2}\;{\rm eV} \;,\; \frac{\varepsilon}{\xi} \sim -0.53 \;,\; \frac{1}{\xi}\sim 6.4\times 10^{-3} \;.
\end{equation}

\ni Through Eqs. (3) the particular mass constraint $m_1/m_2 = 1/\varphi^2$ is equivalent to the following parameter constraint:

\begin{equation}
\frac{9}{4}\; \frac{1-\varepsilon/\xi}{9 - 80/\xi - \varepsilon/\xi} = \frac{m_1}{m_2} = \left( \frac{2}{1 + \sqrt5}\right)^2 = 0.381966
\end{equation}

\ni or

\begin{equation}
\frac{1}{\xi} = 0.0388678 + 0.0611322 \frac{\varepsilon}{\xi}\,,
\end{equation}

\ni where $\varepsilon/\xi$ and $1/\xi$ are estimated by the experiment only up to two decimals (Eq. (22).

Note finally that our original neutrino mass formula (1) consists of three contributions

\begin{equation}
\mu\,\rho_i \;,\; -\mu\,\rho_i\frac{1}{\xi} N^2_i  \;,\;-\mu\,\rho_i\frac{1}{\xi}\; \frac{\varepsilon -1}{N^2_i}\;\;\;(i=1,2,3)\;,
\end{equation}

\ni dependent in three different ways on the numbers $N^2_i = 1,9,25$. The parameter constraint $\varepsilon = 1$ (or $\varepsilon/\xi = 1/\xi$), discussed in this note, eliminates the third contribution, while the parameter constraint $\varepsilon = 0$ (or $\varepsilon/\xi = 0$), considered in the first part of this paper, {\tt hep-ph/0609187}, eliminated the sum of the second and third contributions for the lowest generation $i =1$. The reader may consult Ref. [5] for a possible "intrinsic interpretation"\, of three contributions (25). It is a consistent interpretation based on the idea of algebraically composite fundamental fermions satisfying a generalized Dirac equation (in such an interpretation, $N_1 = 1, N_2 = 3, N_3 = 5$ are the numbers of spin-1/2 algebraic partons involved "within"\, the fundamental fermions, leptons and quarks, of three generations $i=1,2,3$).

Concluding, in this note a two-parameter neutrino mass formula (7) or (8) is described, giving a moderately hierarchical spectrum $m_1 < m_2 < m_3 $, consistent with the experimental estimates of $\Delta m^2_{21}$ and $\Delta m^2_{32} $ treated here as the input. This mass formula follows from a three-parameter empirical neutrino mass formula (1) or (3) through imposing a parameter constraint $\varepsilon = 1$ (or $\varepsilon/\xi = 1/\xi$) and leads to a very simple mass sum rule (9). Some alternative parameter constraints, used tentatively to eliminate one of three parameters, are compared (see our table).

\vspace{0.4cm}

\centerline{*{\hspace{2cm}}*{\hspace{2cm}}*}

\vspace{0.4cm}

In the second half of this note, we would like to report briefly on another approach to the problem of neutrino mass parametrization, starting from the structure of neutrino mass matrix $M = (M_{\alpha\,\beta} )$ related to its eigenvalues $m_i : U^\dagger M U = {\rm diag}(m_1,m_2,m_3)$. To this end assume for the neutrino mixing matrix $U = (U_{\alpha\,i})$ the experimentally favored tribimaximal form [6]

\begin{equation} 
U =   \left( \begin{array}{rrc} \sqrt{\frac{2}{3}} & \frac{1}{\sqrt3} & 0 \\ -\frac{1}{\sqrt6} & \frac{1}{\sqrt3}  & \frac{1}{\sqrt2} \\ \frac{1}{\sqrt6} & -\frac{1}{\sqrt3}  & \frac{1}{\sqrt2} \end{array} \right) \; ,
\end{equation} 

\ni leading to the following mass-matrix elements $M_{\alpha\,\beta} = \sum_i U_{\alpha\,i}m_i U^*_{\beta\,i} $: 

\vspace{-0.2cm}

\begin{eqnarray}
M_{e\,e} & = & \:\:\:\frac{1}{3}(2m_1 + m_2)\,, \nonumber \\
M_{\mu\,\mu} & = & \:\:\,M_{\tau\,\tau} = \:\:\:\,\frac{1}{6} (m_1 + 2m_2 + 3 m_3)\,, \nonumber \\
M_{e\,\mu} & = & -M_{e\,\tau} =  \frac{1}{3} (-m_1 + m_2)\,, \nonumber \\
M_{\mu\,\tau} & = & \frac{1}{6} (-m_1 - 2m_2 + 3 m_3)
\end{eqnarray}

\ni and $M_{\beta\,\alpha} = M_{\alpha\,\beta} $ otherwise. Here, the charged-lepton flavor representation is chosen, where the charged-lepton mass matrix is diagonal and so $U$ is at the same time the lepton mixing matrix (appearing in the charged weak current).

In the tribimaximal case, among nine matrix elements $M_{\alpha\,\beta}$ there are three independent, say, $M_{e\,e}, M_{e\,\mu}, M_{\mu\,\tau}$ , where $M_{\mu\,\mu} = M_{e\,e}  + M_{e\,\mu} + M_{\mu\,\tau}$. It is convenient to introduce here a new parameter $\chi $ equal to the ratio $M_{\mu\,\tau}/M_{e\,\mu}$:

\begin{equation}
M_{\mu\,\tau} = \chi M_{e\,\mu}\;.
\end{equation}

\ni Then, from Eq. (28) and the third and fourth Eqs. (27), we obtain a new neutrino mass sum rule

\begin{equation}
m_3 = \eta m_2 - (\eta - 1) m_1 \;,
\end{equation}

\ni where

\begin{equation}
\eta \equiv \frac{2}{3} (\chi + 1) \;,\; \chi \equiv \frac{3}{2} \eta - 1 \;.  
\end{equation}

\ni This sum rule is parametrized by $\chi$ (the previous sum rule, Eq. (9), corresponds to the parameter constraint (6) and so is parametrically fixed). Of course, the parameter $\chi$ can be easily evaluated, if the mass spectrum is known. This is the case, when in the mass formula (1) or (3) one of our parameter constraints is accepted (see our table) in order to eliminate one parameter.

Now, making use of the mass sum rule (29) and the identity (11), we get for the ratio $r \equiv m_1/m_2$ the quadratic equation

\begin{equation}
[(\eta - 1)^2 + \lambda] r^2 - 2\eta(\eta - 1) r + \eta^2 - \lambda - 1 = 0\;,
\end{equation}

\ni where $\lambda \equiv \Delta m^2_{32}/\Delta m^2_{21} \sim 30$ when the experimental estimates $\Delta m^2_{21} \sim 8.0\times 10^{-5}\;{\rm eV}^2$ and $\Delta m^2_{32} \sim 2.4\times 10^{-3}\;{\rm eV}^2$ are applied as the input. This equation can be factorized into the form

\begin{equation}
(r-1)\left\{[(\eta -1)^2  + \lambda)] r -  \eta^2 + \lambda + 1\right\} = 0 
\end{equation}

\ni giving readily two solutions for $r$ in terms of $\eta$:

\begin{equation}
 r = \left\{ \begin{array}{l} 1 \\ \frac{\eta^2 - \lambda - 1}{(\eta - 1)^2 + \lambda} \end{array}\right. \;.
\end{equation}

The first solution $r = 1$ corresponds to the limiting option of exact neutrino mass degeneracy $m_1 = m_2 = m_3$ which is experimentally excluded. The second solution $r \neq 1$ is nonnegative, $r \geq 0$, {\it only} if $\eta^2 -1 \geq \lambda$, leading then to $m_1 \geq 0$ (we take $m_1, m_2, m_3$ as nonnegative and nondegenerate). For this solution, we infer from Eq. (33) that the parameter $\eta $ satisfies the quadratic equation 

\begin{equation}
(1 - r)\eta^2 + 2 r \eta - (\lambda+1)(1+r) = 0\,,
\end{equation}

\ni implying two solutions for $\eta$ in terms of $r \neq 1$: 

\begin{equation}
\eta = \frac{-r \pm \sqrt{\lambda(1-r^2) + 1}}{1-r} \;.
\end{equation}

\ni Here, $\lambda \sim 30$ when $\Delta m^2_{21} \sim 8.0\times 10^{-5}\;{\rm eV}^2$ and $\Delta m^2_{32} \sim 2.4\times 10^{-3}\;{\rm eV}^2$ are used as the input. 

In the case of parameter constraint $\varepsilon = 1$ (or $\varepsilon/\xi = 1/\xi$) discussed in the present note, we can put in Eq. (35) $r \sim 0.263$ as given in Eq. (15). Then, with $\lambda \sim 30$  the positive solution to Eq. (34) for $\eta$ is

\begin{equation}
\eta \sim 6.94 = 6.9 
\end{equation}

\ni giving {\it via} Eq. (30) the value

\begin{equation}
\chi \sim 9.41 = 9.4 
\end{equation}

\ni (note that $M_{e\,\mu}$ and $M_{\mu\,\tau}$ are positive if $m_1 < m_2 < m_3$ as it holds certainly for $r \sim 0.263$).

In the case, when our other parameter constraint $\varepsilon = \xi $ (or $\varepsilon/\xi = 1$) is considered, we have $m_1 = 0$ and so $r = 0$. Then, with $\lambda \sim 30$  the positive solution to Eq. (34) for $\eta$ becomes

\begin{equation}
\eta = \sqrt{\lambda+1} \sim 5.57 = 5.6 \;,
\end{equation}

\ni leading through Eq. (30) to the value

\begin{equation}
\chi \sim 7.35 = 7.4 \;.
\end{equation}

\ni Notice that here $\chi \sim 1.06\times 4\sqrt3 $. 

In a recent paper [7] we have constructed in the three-generation space of $\nu_1, \nu_2, \nu_3$ an oscillatory model for the off-diagonal part of neutrino mass matrix $M = \left(M_{\alpha\,\beta} \right)$, where it has turned out that $\chi = 4\sqrt3 = 6.92820$ precisely. Then, $\eta = (2/3)(4\sqrt3 + 1) = 5.28547$, implying through Eq. (29) that $\lambda \equiv {\Delta m^2_{32}}/{\Delta m^2_{21}} = \eta^2 - 1 = 26.9362 \stackrel{<}{\sim} 30$ if $m_1 = 0$ (and so $r = 0$). In this case, we {\it predict} $\Delta m^2_{32} \sim 2.2\times 10^{-3}\;{\rm eV}^2$ when $\Delta m^2_{21} \sim 8.0\times 10^{-5}\;{\rm eV}^2$ is applied as the {\it only} input, while the popular experimental estimate is $\Delta m^2_{32} \sim 2.4\times 10^{-3}\;{\rm eV}^2$ giving $\lambda \sim 30$. Here, $m_2 = \sqrt{\Delta m^2_{21}} \sim 8.9\times 10^{-3}\;{\rm eV}$ and $m_3 \equiv \sqrt{\Delta m^2_{32}+ m^2_2} = m_2\sqrt{\lambda+1} \sim 4.7\times 10^{-2}$ eV if $m_1 = 0$. Hence, due to Eqs. (5), $\mu \sim 8.1\times 10^{-2}$ eV, $\varepsilon/\xi =1$ and $1/\xi = 1,03311\times 10^{-2}$ if $m_1 = 0$ (the precise value of $1/\xi $ follows from $m_3 = \lambda m_2$ and the cancellation of $m_2$). If $m_1 >0$ (and so $r > 0$), then $\lambda < \eta^2 - 1 = 26.9362 $  since $\eta^2- \lambda - 1 = [(\eta-1)^2 + \lambda] r > 0$ due to the second Eq. (29). Thus, in this case, the value $m_1 > 0$ has the tendency of spoiling the approximate agreement of $\lambda$ with its experimental estimate $\lambda \sim 30$. Note that for $\chi = 4 \sqrt3 $ the neutrino mass sum rule (29) is equivalent to the parameter constraint

\begin{equation}
\frac{1}{\xi} = 0.017363 - 0.0070452 \frac{\varepsilon}{\xi}\,.
\end{equation}

\ni when Eqs. (3) are used. This gives $1/\xi = 1.03311\times 10^{-2}$ if $m_1 = 0$ ({\it i.e.}, $\varepsilon/\xi = 1$).

More generally, if -- as in Ref. [7] -- a value of $\chi $ is provided by  {\it another} argument than our mass formula (1) or (3) jointly with a parameter constraint (see our table), and if  $m_1 = 0$ (and so $r = 0$), then $\lambda = \eta^2 - 1$ with $\eta \equiv (2/3)(\chi + 1)$ is implied by the second Eq. (33) (following from our new mass sum rule (29)). In this case, we {\it predict} $\Delta m^2_{32} \equiv \lambda \Delta m^2_{21} $, where the experimental input of {\it only} $\Delta m^2_{21} \sim 8.0\times 10^{-5}\;{\rm eV}^2$ is applied. Here, $m_2 =\sqrt{\Delta m^2_{21}} \sim 8.9\times 10^{-3}\;{\rm eV}$ and $m_3 \equiv \sqrt{\Delta m^2_{32} + m^2_2} = m_2 \sqrt{\lambda+1} $ if $m_1 = 0$. In order to get $\lambda \sim 30$ in the case of $m_1 = 0$ one ought to have $\eta = \sqrt{\lambda +1} \sim 5.6$ and $\chi = (3/2)\eta - 1 \sim 7.4$, as in Eqs. (38) and (39) valid for our parameter constraint $\varepsilon = \xi$ (or $\varepsilon/\xi = 1$). If the mass $m_1 > 0$ (and so the ratio $r > 0$), then $\lambda < \eta^2 - 1 $ because of the second Eq. (33) that implies $\eta^2 - \lambda -1 = [(\eta - 1)^2 +\lambda] r > 0$.

\vfill\eject

~~~~
\vspace{0.5cm}

{\centerline{\bf References}}

\vspace{0.5cm}

{\everypar={\hangindent=0.6truecm}
\parindent=0pt\frenchspacing

{\everypar={\hangindent=0.6truecm}
\parindent=0pt\frenchspacing

\vspace{0.2cm}

[1]~W. Kr\'{o}likowski, {\tt hep--ph/0610248}.

\vspace{0.2cm}

[2]~W. Kr\'{o}likowski, {\it Acta Phys. Polon.} {\bf B 37}, 2601 (2006) [{\tt hep--ph/0602018}].

\vspace{0.2cm}

[3]~{\it Cf. e.g.} G.L. Fogli, E. Lisi, A. Marrone, A. Palazzo, {\it Progr. Part. Nucl. Phys.} {\bf 57}, 742 (2006) [{\tt hep--ph/0506083}]; M.C. Gonzalez-Garcia, M. Maltone, {\tt ar$\chi$iv: 0704.1800 [hep--ph]}.

\vspace{0.2cm}

[4]~Y. Kajiyama, M. Raidal, A.Strumia, {\tt ar$\chi$iv: 0705.4559 [hep--ph]}.

\vspace{0.2cm}

[5]~W. Kr\'{o}likowski, {\tt hep--ph/0604148}; and references therein.

\vspace{0.2cm}

[6]~L. Wolfenstein,  {\it Phys. Rev.} {\bf D 18}, 958 (1978); P.F. Harrison, D.H. Perkins, W.G.~Scott, {\it Phys. Lett.} {\bf B 458}, 79 (1999); {\it Phys. Lett.} {\bf B 530}, 167 (2002); Z.Z.~Xing. {\it Phys. Lett.} {\bf B 533}, 85 (2002); P.F. Harrison, W.G.~Scott, {\it Phys. Lett.} {\bf B 535}, 163 (2003); T.D.~Lee, {\tt hep--ph/0605017}.  

\vspace{0.2cm}

[7]~W. Kr\'{o}likowski, {\it Acta Phys. Polon.} {\bf B 37}, 2805 (2006) [{\tt hep--ph/0606223}].

\vspace{0.2cm}
\vspace{0.2cm}

\vfill\eject

\end{document}